\documentclass[aps,pre,preprint]{revtex4}
\usepackage{graphicx,epsfig}
\begin{document}
\title{Information feedback and mass media effects in cultural dynamics}
\author{J.C. Gonz\'alez-Avella}
\affiliation{IMEDEA (CSIC-UIB), Campus Universitat Illes Balears,
E-07122 Palma de Mallorca, Spain}
\author{M. G. Cosenza}
\affiliation{Centro de F\'isica Fundamental, Universidad de los
Andes, M\'erida, M\'erida 5251, Venezuela}
\author{K. Klemm}
\affiliation{Bioinformatics, Department of Computer Science,
University of Leipzig,  04107 Leipzig, Germany}
\author{V. M. Egu\'iluz}
\affiliation{IMEDEA (CSIC-UIB), Campus Universitat Illes Balears,
E-07122 Palma de Mallorca, Spain}
\author{M. San Miguel}
\affiliation{IMEDEA (CSIC-UIB), Campus Universitat Illes Balears,
E-07122 Palma de Mallorca, Spain}
\date{  }
\begin{abstract}
We study the effects of different forms of information feedback associated with mass media on an agent-agent based model of the dynamics of cultural dissemination. In addition to some processes previously considered, we  also examine  a model of local mass media influence in cultural dynamics. Two mechanisms of information feedback are investigated: (i) direct mass media influence, where local or global mass media act as an additional element in the network of interactions of each agent, and (ii) indirect mass media influence, where global media acts as a filter of the influence of the existing network of interactions of each agent. Our results generalize previous findings showing that cultural diversity builds-up by increasing the strength of the mass media influence. We find that this occurs independently of the mechanisms of action (direct or indirect) of the mass media message. However, through an analysis of the full range of parameters measuring cultural diversity, we establish that the enhancement of cultural diversity produced by interaction with mass media only occurs for strong enough mass media messages. In comparison with previous studies a main different result is that weak mass media messages, in combination with agent-agent interaction, are efficient in producing cultural homogeneity. Moreover, the homogenizing effect of weak mass media messages are more efficient for direct local mass media messages than for global mass media messages or indirect global mass media influences. 
\end{abstract}
%\pacs{ }
\maketitle
\section{Introduction}
In an influential paper, Robert Axelrod \cite{Axelrod} addressed the
question
\begin{quote}
if people tend to become more alike in their beliefs, attitudes,
and behavior when they interact, why do not all differences
eventually disappear?.
\end{quote}
To investigate this problem, Axelrod introduced an agent-based
model to explore mechanisms of competition between the tendency
towards globalization and the persistence of cultural diversity.
Culture in this model is defined as a set of individual attributes
subject to social influence.  The state of an agent is described
by a set of $F$ cultural features, each with $q$ possible values
or traits. The parameter $q$ gives a measure of initial culture
diversity for a random distribution of trait values. The local
interaction between neighboring agents follows two basic social
principles. The probability of interaction is determined by a
principle of homophily: it is proportional to the cultural overlap
between the agents, that is to the amount of cultural similarities
(number of features) that they share. The result of the
interaction is determined by a principle of social influence, so
that similarity is enhanced when interaction occurs. In a typical
dynamical evolution, the system freezes in a multicultural state
with coexisting spatial domains of different cultures,
illustrating how a simple mechanism of local convergence can lead
to global polarization. Later systematic studies of Axelrod's
model have identified a globalization-polarization transition
depending on the value of $q$ for a fixed $F$
\cite{Castellano,Maxi2,MaxiR,Maxi4,JEDC}. There is a threshold
value $q_c$ such that for $q<q_c$ globalization (cultural
homogeneity) occurs, while for $q>q_c$ a persistent multicultural
state (polarization or cultural diversity) is reached.

Several extensions of this model have been investigated, some of
them already suggested in Axelrod's paper. For example, it has
been shown that frozen polarized configurations are not robust
against cultural drift modelled as random perturbations
\cite{Maxi2}. Other extensions include the consideration of
quantitative instead of qualitative values for the cultural traits
\cite{Flache1}, the extension of the model to continuous values of
the cultural traits and the inclusion of heterophobic interactions
\cite{Flache2}, the simulation of technology assimilation
\cite{JASSS1}, the consideration of specific historical contexts
\cite{JASSS2}, or the effect of a fixed external cultural
influence \cite{GCT}. A number of works have addressed the issue
of the role of the social network of interactions. The existence
of long range links of interaction in a social network with a
small world structure naturally promotes cultural globalization
\cite{Maxi1}. However, a study of the increase in the range of
interaction \cite{Greig} suggests that an increase in
communication promotes the emergence of a global culture, but it
can also function to maintain areas of cultural uniqueness. A
related question is the co-evolution of the social network of
interaction and the dynamics of cultural changes, so that possible
social interaction are not fixed from the outset \cite{Damon}.
Within this general context of the studies of the effects of
different forms of social interactions, Shibanai et
al.\cite{Shibanai} considered the process of information feedback
into the social system. Such feedback mechanism is one of the
functions of mass media. From a general perspective, this question
can be addressed as the competition of local agent interactions
with mechanisms of global coupling \cite{PRE}. Shibanai et al.
concluded that
\begin{quote}
mass media, contrary to lay beliefs of their
strong uniforming power, would rather contribute to creating
differences in the long run.
\end{quote}
This result is certainly surprising since mass media are believed
to be  powerful instruments to influence people's attitudes and
opinions to homogenize society.
Conclusions in reference \cite{Shibanai} are based on the analysis
for a single value of the parameter $q$ and few values of the intensity of the influence of mass media. In this paper we consider
the full range of values of $q$ when addressing the general
question of the effects of different forms of information feedback
on cultural dynamics in the framework of Axelrod's model, including the study of local feedback. We
consider information feedback mechanisms as different types of
mass media influences acting on a social system. This extension of
Axelrod's model was referred to as ``public education and
broadcasting" \cite{Axelrod}. Our aim is to explore mechanisms,
and their efficiency, by which mass media modifies processes of
cultural dynamics based on local agent interactions. We
investigate two main mechanisms that differ on how the information
feedback affects the agents \cite{Shibanai}: (i) direct mass media
influence, and (ii) indirect mass media influence. In the first
case, mass media acts as an additional element in the network of
interactions of each agent. As applications of direct mass media
influence we study models of global and local mass media and
compare their effects on the system. In the second feedback
mechanism, mass media acts as a moderator or filter of the
influence of the existing network of interactions of each agent.
The assumption of this second model is that mass media have
indirect influential power which reinforces the effect of personal
networks. In particular, we focus on a model of indirect influence
of global mass media.

We deal with states of the agents and mass media influences
described by vectors whose components can take discrete values. We
consider mass media influences that originate endogenously but can
act either directly or indirectly on the system. In the first
case, the agent-agent interaction and the interaction of the
agents with the mass media are two independent processes and both
are based on the same homophily and social influence principles of
Axelrod's model. This scheme constitutes a model for a social
system interacting with global or local mass media that represent
endogenous cultural influences or plurality information feedback
at different levels. In the case of indirect action,  agent-agent
interactions is not independent of the state of global mass media,
the interaction being reinforced when the states of the agent and
mass media message are connected.

Our results show that the conclusions of Shibanai et al. \cite{Shibanai} are not generally valid, but that different effects are obtained for different values of the parameters of the system.  We find that cultural diversity builds-up by increasing the
probability of interaction of the agents with the mass media, in agreement with Shibanai et al. This
occurs independently of the form of action (either direct or
indirect) of the mass media message. However, a new finding that follows from our simulations is that
cultural diversity is only promoted by strong enough mass media
messages, while weak mass media messages, in combination with
agent-agent interaction, are efficient in producing cultural
homogenization. Moreover, this homogenizing effect in the way to
globalization turns out to be much more efficiently implemented by
direct local mass media messages, as for example local or regional
TV (narrowcast) than by direct global mass media messages as
worldwide TV channels (broadcast), or by an indirect influence of
global mass media as in a filtering process of local interactions.

The model of direct mass media influence, including the
description of the local and global mass media, is presented in
Section~II.  This section shows the effects of the direct action
of these two types of mass media. We describe different effects
occurring in the culturally homogeneous state of the system
($q<q_c$) as well as in the multicultural state ($q>q_c$).
Section~III describes the model of indirect global mass media
influence and the effects of this mechanism in the collective
states of the agents. Section~IV contains the conclusions of our
work, including a comparison of the effects of the different
models of information feedback that we have considered.

\section{Models of direct mass media influence}
We consider a system of $N$ agents at the nodes of a
two-dimensional regular lattice. The state $c_i$ of agent $i$ is
defined by a vector of $F$ components $\sigma_i=
(\sigma_{i1},\sigma_{i2},\ldots,\sigma_{iF})$. In Axelrod's model,
the $F$ components of $c_i$ correspond to the cultural features
(language, religion, etc.) describing the $F$-dimensional culture
of agent $i$. Each component $\sigma_{if}$ can take any of the $q$
values in the set $\{ 0, 1, \ldots, q-1 \}$  (called cultural
traits in Axelrod's model). As an initial condition, each agent is
randomly and independently assigned one of the $q^F$ possible
state vectors with uniform probability. We introduce a vector $M$
representing the mass media message, with components
$(\mu_{i1},\mu_{i2},\ldots,\mu_{iF})$. We distinguish two types of
direct mass media influences, schematically shown in Fig.~1.

\begin{figure}[h]
\includegraphics[width=0.4\linewidth,angle=0]{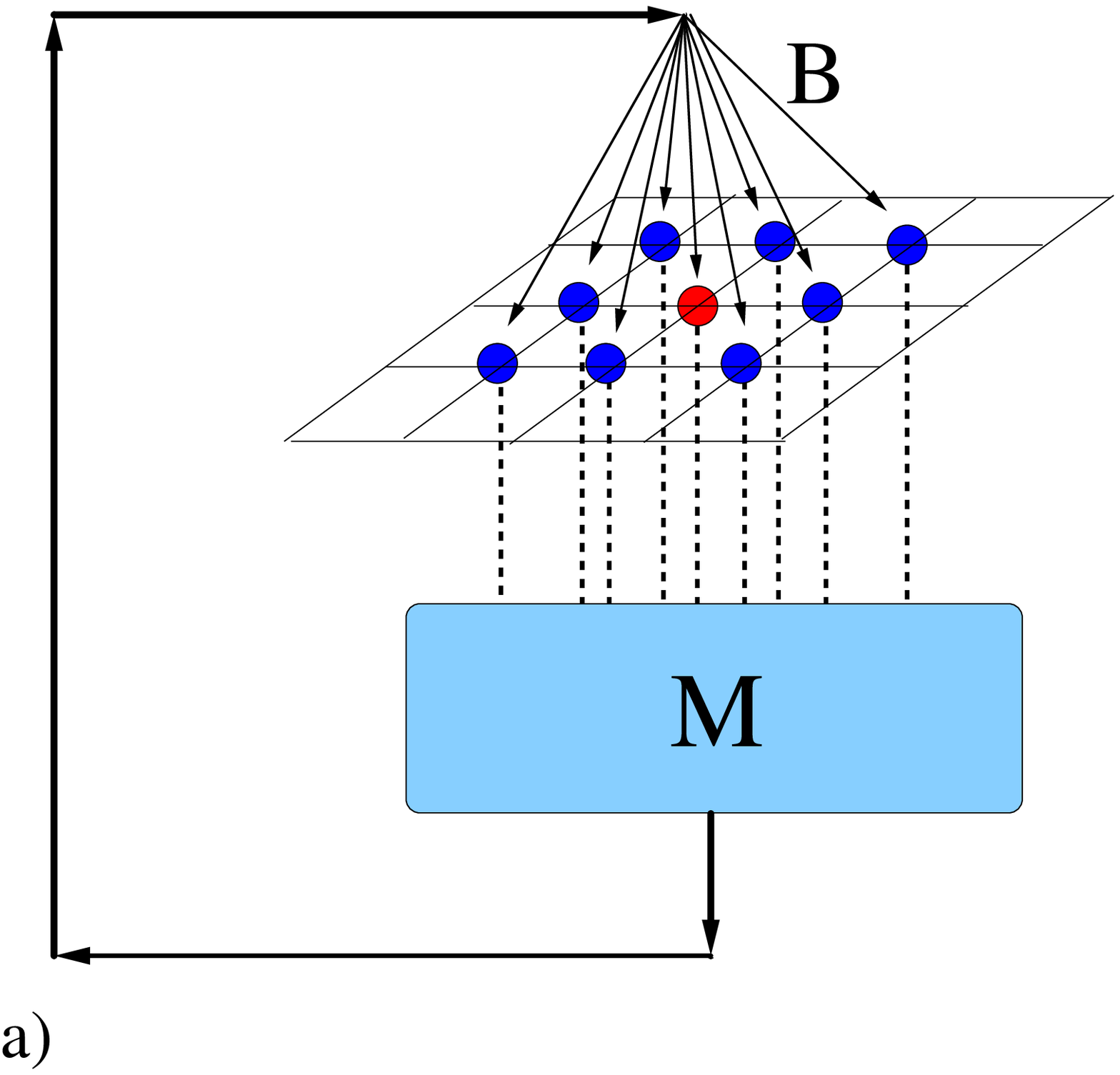}
\includegraphics[width=0.4\linewidth,angle=0]{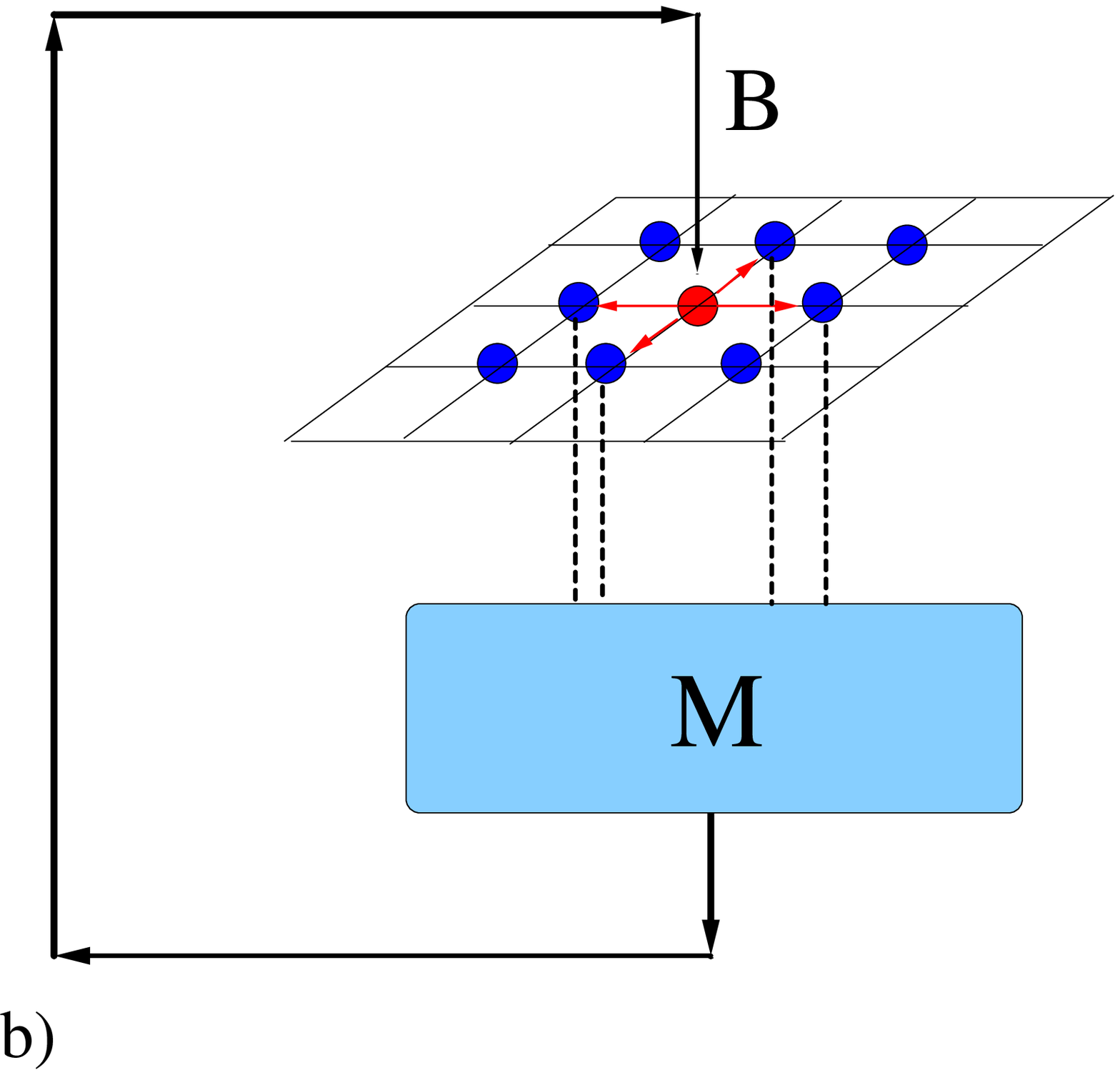}
\caption{Diagrams representing two types of direct, endogenous mass media
influences acting on the system.
a) Global mass media. b) Local mass media.}
\end{figure}

(i) The \emph{global mass media} is spatially uniform and may vary
in time. Each $\mu_{if}$ is assigned the most abundant value
exhibited by the $f$-th component of the state vectors of al the
agents, which we denote by $\mu_f$. If the maximally abundant
value is not unique, one of the possibilities is chosen at random
with equal probability. In this type of influence the same global
information is feedback to each agent at any given time. The
components of $M$ may change  as the system evolves. In the
context of cultural models, this represents a global mass media
influence shared identically by all the agents and containing the
most predominant  trait in each cultural feature present in a
society (a ``global cultural trend") \cite{Shibanai}.

(ii) The \emph{local mass media}, is spatially non-uniform and
non-constant on time. Here the component $\mu_{if}$ is assigned
the most frequent value of the $f$-th component of the state
vectors of the agents belonging to the von Neumann neighborhood of
agent $i$. If there are two or more maximally abundant values of
component $f$ one of these is chosen at random with equal
probability. This type of influence can be interpreted as a local
mass media conveying the ``local cultural trend" of its
neighborhood to each agent in a social system.

Systems subject to either local or global mass media describe
social systems with endogenous cultural influences. Cultural
influences generated endogenously represent a plurality
information feedback, which is one of the functions of mass media
\cite{Shibanai}, but this can occur at a global (''broadcast") or
at a local (''narrowcast") level.

The strength of the coupling to the mass media is controlled by a
parameter $B \in [0,1]$ that measures the probability of
interaction of the agents with the mass media message. We shall
assume that $B$ is uniform, i.e., the mass media influence reaches
all the agents with the same probability. The parameter $B$ can be
interpreted as the probability that the mass media message
attracts the attention of the agents in the social system. The
parameter $B$ takes into account factors of the mass media
influence that can be varied, such as its amplitude, frequency,
attractiveness, etc. At any given time, we assume that any agent
can either interact with the mass media message or with other
agents in the system. Each agent in the network possesses a
probability $B$ of interacting with the message $M$ and a
probability $(1-B)$ of interacting with its neighbors.

Formally, we treat the mass media acting on each agent $i$ as an
additional neighbor of $i$ with whom an interaction is possible.
The mass media message is represented as an additional agent
$\phi(i)$ such that $\sigma_{\phi(i) f} = \mu_{if}$ in the
following
definition of the dynamics. The dynamical evolution proceeds iterating the following steps: \\

(1) Select at random an agent $i$ on the lattice (called active
agent).

(2) Select the source of interaction $j$. With probability $B$ set
$j=\phi(i)$ as an interaction with the mass media vector. Otherwise, choose
agent $j$ at random among the four nearest neighbors of $i$ on the network.

(3) Calculate the cultural overlap (number of shared features)
$l(i,j)=\sum_{f=1}^{F}\delta_{\sigma_{if},\sigma_{jf}}$. If $0<l(i,j)<F$,
sites $i$ and $j$ interact with probability $l(i,j)/F$. In case of interaction,
choose $h$ randomly such that $\sigma_{ih}\neq \sigma_{jh}$ and set
$\sigma_{ih} = \sigma_{jh}$. Here we use the definition of the Kroneckr's delta function.

$
 \delta_{\sigma{if},\sigma_{jf}}\left\{
      \begin{array}{c}
        if \; \; \sigma_{if}= \sigma_{jf}; \; \; \delta_{\sigma{if},\sigma_{jf}} = 1\\
        if \; \; \sigma_{if} \neq \sigma_{jf}; \; \; \delta_{\sigma{if},\sigma_{jf}} = 0
      \end{array}. \right.
$
\\

(4) Update the mass media vector $M$ if required. Resume at (1). \\

The role of mass media can be seen as an additional neighbor of
each agent $i$ with whom an interaction can take place. This
``fifth" neighbor competes with the other four neighbors to
influence an active agent $i$. $B$ is the probability of capturing
the attention of any agent to interact with either the global or
the local mass media message $M$. The effect of that interaction
on each agent depends on the specific cultural overlap between the
agent and the message $M$. The total probability to interact with
the mass media is $B p_{iM}$, where $p_{iM}=l(i,M)/F$, and
$l(i,M)$ is the overlap between agent $i$ and the message. On the
other hand, the active agent $i$ has a probability
$\frac{1}{4}(1-B) p_{ij}$ to interact with a randomly chosen
neighbor $j$. Therefore, the parameter $B$ weights the influence
of the mass media with respect to the influence of the neighbors
on an agent $i$. This dynamics is based on the same principles of
Axelrod's model, namely homophily or similarity and social
influence. But now the agents interact either between them or with
M. The competition of these interactions in measured by $B$.

To characterize the collective final state reached by the
dynamics, we calculate the average fraction of cultural domains
$g=\langle N_g \rangle /N$. A cultural domain is a set of
contiguous sites with identical cultural traits. $N_g$ is the
number of cultural domains formed in the final state of the
dynamics for a given realization of initial conditions. A
culturally homogeneous state is characterized by values $g
\rightarrow 0$. When the system settles into a culturally diverse
state we have $\langle N_g \rangle \gg 1$. Our numerical results
in this work are based on averages over $50$ realizations for
systems of size $N=50\times 50$, and $F=5$.

Figure~2 shows the average fraction of cultural domains $g$ as a
function of time under the direct action of global mass media, for
two values of the number of traits $q$ with $F=5$, and for
different values of the probability $B$. Fig.~2 (left): for small
values of $q$ the system reaches a culturally homogeneous state in
absence of mass media influence ($B=0$) and also for small values
of $B$. When the probability $B$ increases, we see that global
mass media is able to induce cultural diversity in the system.
This result agrees with the results obtained by Shibanai et al.
\cite{Shibanai} about the ability of global mass media to induce
cultural diversity in the system ($g\neq 0$). However we observe
in Fig.~2 (right) that, for larger values of $q$, where the system
would be in a culturally polarized state with no mass media
influence ($B=0$), small values of the probability $B$ can reduce
the number of cultural groups. Effects similar to those shown in
Fig.~2 are observed when a local mass media message interacts with
the system. We therefore conclude that mass media can increase or
decrease cultural diversity depending on the value of the
parameter $q$.

\begin{figure}[h]
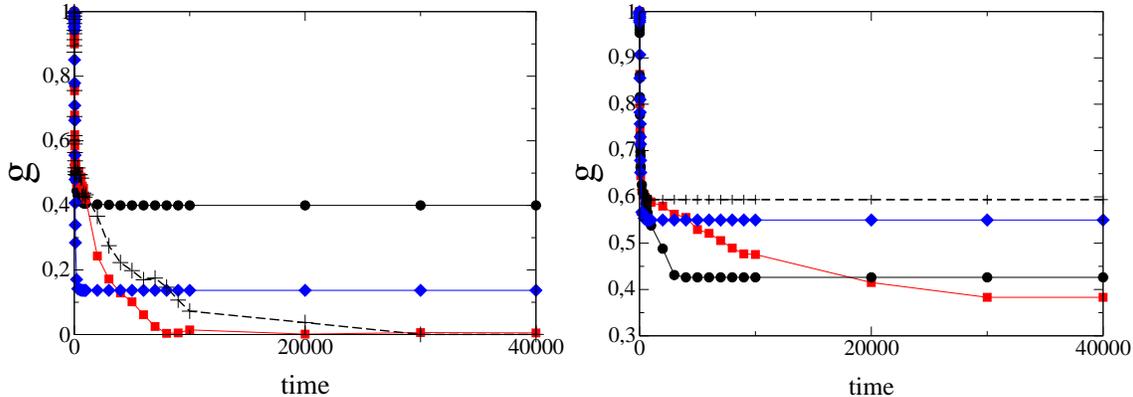

\includegraphics[width=0.45\linewidth,angle=0]{Fig2a.eps}
\includegraphics[width=0.45\linewidth,angle=0]{Fig2b.eps}
\caption{Evolution of $g$ in a system subject to a global mass
media message for different values of the probability $B$, with
fixed $F=5$. Time is measured in number of events per site. System
size $N=50 \times 50$. Left: $q=10$; $B=0$ (crosses); $B=0.0005$
(squares); $B=0.15$ (diamonds); $B=0.6$ (circles). Right: $q=30$;
$B=0$ (crosses); $B=0.0005$ (squares); $B=0.005$ (circles);
$B=0.1$ (diamonds).}
\end{figure}

A global picture of direct mass media influences is provided by
Figure~3, which shows the asymptotic value of $g$ for long time as
a function of $q$, with $F=5$, for different values of the
probability $B$. In absence of any mass media influences ($B=0$)
there is a threshold value of the number traits $q_c \approx 25$,
such that for $q<q_c$ dynamical evolution always leads to one of
the possible $q^F$ homogeneous states $(g\rightarrow0)$. On the
other hand, for values of $q > q_c$, the behavior of the system
changes and it settles into a multicultural state when $B=0$. When
the probability $B$ is increased, the threshold value of $q$
decreases. There is a value $q_c(B)$ below which the system still
reaches in its evolution a homogeneous cultural state
($g\rightarrow 0$) under the action of any of these mass media
messages.

\begin{figure}[h]
\includegraphics[width=0.5\linewidth,angle=0]{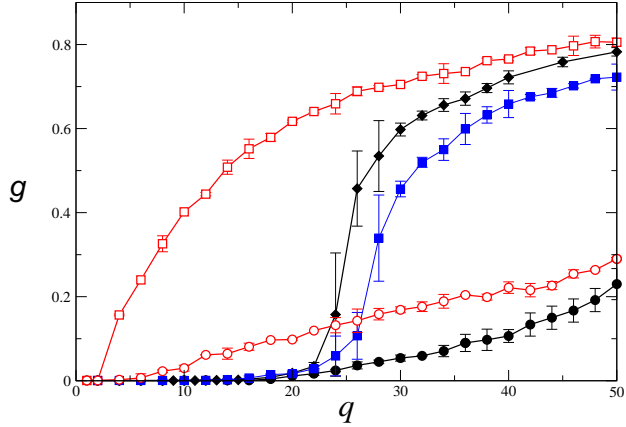}
\caption{Asymptotic value of the fraction of cultural domains $g$
as a function of $q$, for different values of the probability $B$
and for different types of mass media influences. $B=0$
(diamonds); $B=0.0005$ (solid squares, direct global mass media);
$B=0.6$ (empty squares, direct global mass media); $B=0.0005$
(solid circles, direct local mass media); $B=0.1$ (empty circles,
direct local mass media).}
\end{figure}

 A color code can be assigned to each cultural state for visualization of the spatial patters during simulations, as shown in Figure 10 in the Appendix. Figure~4 shows the spatial configurations of the final states of
the system subject to a global mass media message, when $q < q_c$.
In the absence of mass media influence, i.e.  $B=0$, the system
settles into any of the possible $q^F$  homogeneous cultural
states. When the probability $B$ is increased, the system is
driven towards a homogeneous cultural state that depends on the
evolution of the global mass media from the initial conditions.
Thus, for small values of $B$, global mass media contributes to
maintain a globalized state in the system, as one may expect.
However, there is a value of the strength $B$ above which there is
no convergence to a homogeneous state, but a state of cultural
diversity emerges in the system. A similar behavior is observed
when the system is under the influence of local mass media. These
results describe for $q<q_c$ a transition at a threshold value of
$B$ from a culturally homogeneous state to a state of cultural
diversity characterized by an increasing number of cultural
domains as $B$ is increased. Thus, we find the counterintuitive
result that for $q <q_c$, above some threshold value of the
probability of interaction, mass media induces cultural diversity
in a situation in which the system would be culturally homogeneous
under the effect alone of local interactions among the agents.

\begin{figure}[h]
\includegraphics[width=0.5\textwidth]{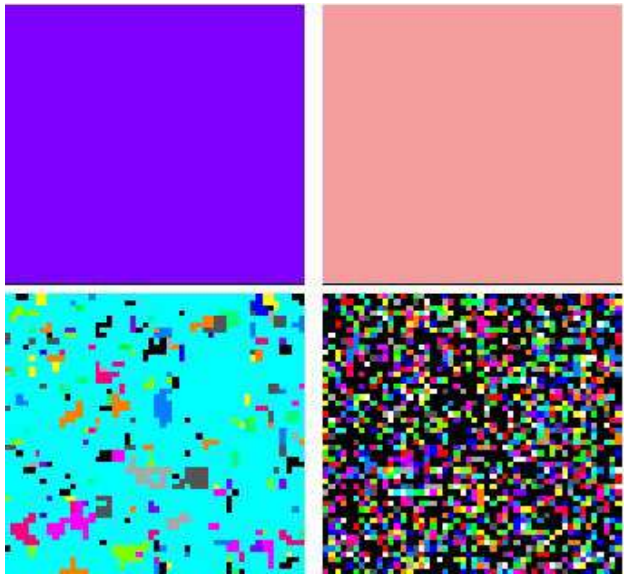}
\caption{Asymptotic cultural configurations for different values
of the probability $B$ for a direct global mass media influence,
for $F=5$, $q=10$, and $N= 50 \times 50$. Top left: $B=0$; top
right: $B=0.01$; bottom left:  $B=0.1$; bottom right: $B=0.9$.}
\end{figure}

Figure~5 shows $q_c$ as a function of $B$. The threshold value
$q_c$ for each type of media decreases with increasing $q$ for
$q<q_c$. In each case, the threshold curve $q_c$ versus $B$ in
Fig.~4 separates the region of cultural diversity from the region
where homogeneous, monocultural states occur on the space of
parameters $(B,q)$. Figure 5 can also be understood as a threshold curve of $B$ versus $q$. For each value of $q$ there is a critical value $B_c(q)$, such that for $B<B_c(q)$, a homogeneous state in reached by the system while for $B>B_c(q)$ multiculturality sets in. For values of $B$ above this curve, the
interaction with the mass media dominates over the local
interactions among the individual agents in the system.
Consequently, agents whose states exhibit a greater overlap with
the state of the mass media have more probability to converge to
that state. This process contributes to the differentiation of
states between neighboring agents and to the formation of multiple
cultural domains in the system for large enough values of the
probability $B$.

\begin{figure}[h]
\includegraphics[width=0.5\textwidth]{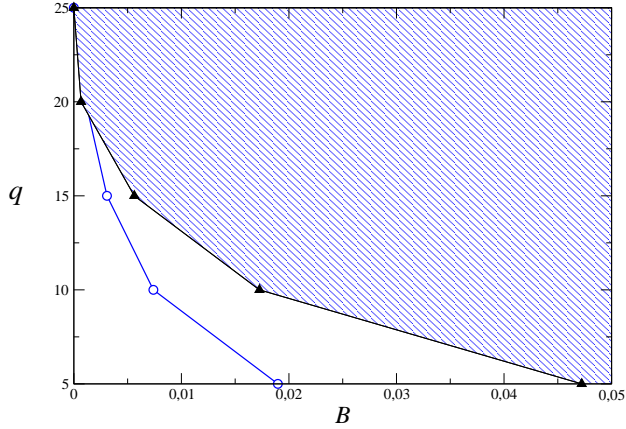}
\caption{Threshold boundaries $q_c(B)$ vs. $B$ for $q<q_c$
corresponding to the global and mass media. Each line separates
the region of cultural diversity (above the line, in grey)  from
the region of a global culture (below the line) for direct global
(circles) and local (triangles) mass media influences.}
\end{figure}

When there are no mass media influences ($B=0$), the system always
freezes into culturally polarized states for $q>q_c$. Figure~3
shows that the effect of mass media for $q>q_c$ depends on the
magnitude of $B$. For the two types of mass media messages that we
consider, small values of $B$ produce a drop in $g$ to values
below the reference line corresponding to its value when $B=0$.
Thus, the limit $B\rightarrow 0$ does not recover the behavior of
the model with only local agent-agent interactions. The fact that
for small values of $B$ the interaction with mass media promotes
cultural homogeneity is related to the non-stable nature of the
inhomogeneous states in Axelrod's model. When the probability of
interaction $B$ is very small, the action of mass media can be
seen as a sufficient perturbation that allows the system to escape
from polarized states with frozen dynamics. The role of mass media
in this situation is similar to that of cultural drift
\cite{Maxi2}.

The drop in the asymptotic values of $g$ for small values of $B$
from the reference value ($B=0$) that is observed for the local
mass media in Fig.~3 is much more pronounced than the
corresponding drops for global mass media. This can be understood
in terms of a greater efficiency of a local mass media influence
as a nonuniform perturbation that allows the system to escape from
a frozen inhomogeneous configuration. Increasing the value of $B$
results, in both types of mass media, in an enhancement of
cultural polarization in the system, but the local mass media
always keeps the amount of cultural diversity, as measured by $g$,
below the value obtained for $B=0$. Information feedback at the
regional level is more efficient in promoting cultural
homogenization. Figure~6 shows configurations reached under the
influence of the direct local and global mass media in the
multicultural region $q > q_c$. A smaller number of cultural
domains (smaller $g$) that for $B=0$ are observed for  global mass
media field. But this number in much smaller under the action of
local mass media field. In addition, the time to reach the final
state is much longer for a local than for a global field (see
Movie 1). Local mass media does not lead to an early frozen state,
leaving room for agent-agent interactions that result in a
culturally homogenized state.

\begin{figure}[h]
\centerline{
\includegraphics[width=0.6\textwidth]{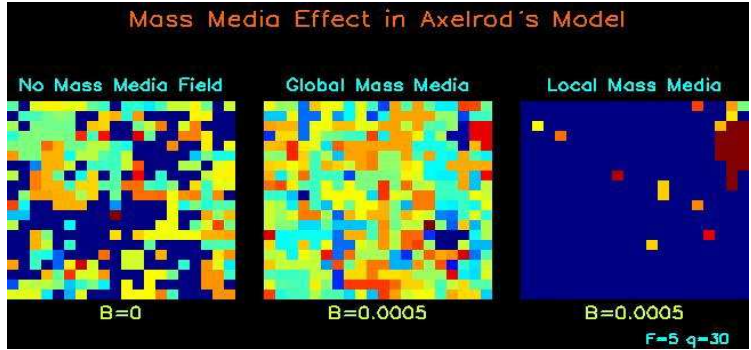}}
\caption{Cultural configurations for different values of the probability
$B$ for different mass media influences in the multicultural region,
for $F=5$, $q=30$, and $N= 50 \times 50$. Left: no mass media $B=0$; center:
Global mass media with $B=0.0005$; right:  Local mass media with $B=0.0005$.}
\end{figure}

We have focused on the effects of information feedback when both the probability of interaction with mass media B and the number of options q  are varied, since they give rise to different qualitative behavior. Variations of other parameters, such as the number of neighbours or the system size, have little effect. In particular, the system size may become relevant only in the limit of very small values of B  for $q < qc$ , when the role of mass media is similar to that of noise added to the system \cite{Maxi2}.

\section{Model of indirect global mass media influence: the filtering of local interactions}

In this section we analyze a model of global information feedback
where the global mass media acts as a moderator or filter of the
local influence of neighbors, as proposed by Shibanai et al.
\cite{Shibanai}. In the original Axelrod's model one feature with
different traits for two neighboring agents is chosen, and the
trait of the active agent is changed to that of the neighbor. This
is modified in the model of indirect global mass media influence
analyzed here, taking into account the agreement of the chosen
trait of the neighbor and that of the global mass media or {\it
the plurality} of the population. If the trait of the neighbor is
concordant with the dominant one, that is, the same as that of the
global mass media message $M$, the feature of the active agent
will be changed to that of the neighbor. But if the feature of the
neighbor is different from that of the global mass media message
$M$, then, with probability $R$ the active agent will not change.
Thus, this model assumes that agents are more likely to adopt a
trait from those neighbors that are concordant with the plurality.

We use the definition of a uniform global mass media as in Section
II, $M=(\mu_{i1},\mu_{i2},\ldots,\mu_{iF})$. The dynamical
evolution of the filter model can be described in terms of the
following
iterative steps: \\

(1) Select at random an agent $i$ on the lattice (active
agent).

(2) Select at random one agent $j$ among the four neighbors of $i$.

(3) Calculate the overlap $l(i,j)$. If $0<l(i,j)<F$, sites $i$ and
$j$ interact with probability $p_{ij}=l(i,j)/F$. In case of
interaction, choose $h$ randomly such that $\sigma_{ih}\neq
\sigma_{jh}$. If $\sigma_{jh}=\mu_{h}$, then set
$\sigma_{ih}=\sigma_{jh}$; otherwise with probability $R$ the
state of agent $i$ does not change and with probability $1-R$ set
$\sigma_{ih}=\sigma_{jh}$

(4) Update the global mass media vector $M$ if required. Resume at (1). \\

Figure~7 shows a diagram of the filter model. The parameter $R$
describes the intensity of the filtering effect of the global mass
media on the local interactions. The case $R=0$ corresponds to the
original Axelrod's model, while $R=1$ implies that cultural
interaction only causes a change if the chosen trait of the
neighbor was equal to that of the global mass media. The overall
probability of interaction between an active agent $i$ and a
chosen neighbor $j$ is $p_{ij} (1-R)$ if the chosen trait of $j$
is different from that corresponding to $M$, and $p_{ij}$ if the
chosen trait is equal to that corresponding to $M$.

\begin{figure}[h]
\centerline{
\includegraphics[width=0.5\textwidth]{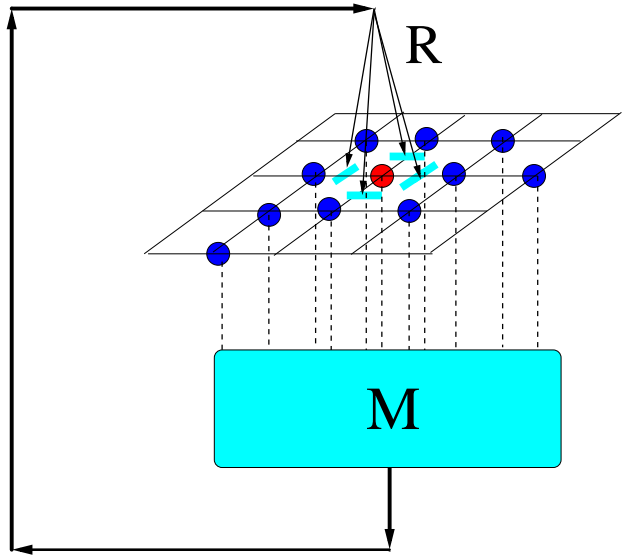}}
\caption{Diagram representing the filter model.}
\end{figure}

Figure~8 shows the average fraction of cultural domains $g$ as a
function of time in the global mass media filter model, for two
values of $q$ with $F=5$, and for different values of the
filtering probability $R$. In Fig.~8 (left), when $q<q_c$ the
system reaches a homogeneous state for $R=0$ and also for small
values of $R$. However, when the probability $R$ increases, the
filtering influence of the global mass media can induce cultural
diversity. Our results for $q<q_c$ support the results obtained by
Shibanai et al. \cite{Shibanai} about the ability of the filtering
process to induce cultural diversity in the same fashion as the
model with direct global mass media influence. But comparison with
Fig.~2 (left) shows that direct interaction with global mass media
is more efficient in promoting cultural diversity than the
filtering mechanism of agreement with the global plurality. The
analysis of reference \cite{Shibanai} was restricted to a single
value of $q<q_c$. We have also explored values of $q>q_c$, where
the system would be in a heterogeneous cultural state in absence
of any filtering ($R=0$). For these values of $q$ we find (Fig.~8,
right) that the filtering mechanism has no appreciable effects for
small $R$, in contrast with the case of direct global mass media
influence where for small values of the probability of interaction
$B$ with the media message, the number of cultural groups is
reduced as a consequence of this interaction.

\begin{figure}[h]
\centerline{
\includegraphics[width=0.75\textwidth]{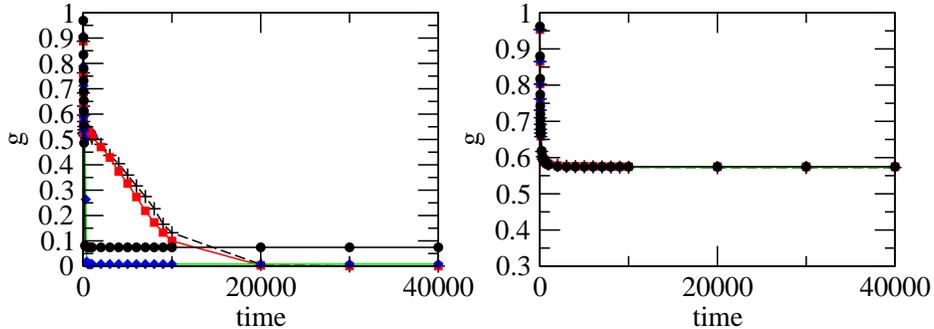}}
\caption{Time evolution of the average fraction of cultural
domains $g$ in the filter model for different values of the
probability $R$, with fixed $F=5$. Time is measured in number of
events per site. System size $N=50 \times 50$. Left: $q=10$; $R=0$
(crosses); $R=0.0005$ (squares); $R=0.15$ (diamonds); $R=0.6$
(circles). Right: $q=30$; $R=0$ (crosses); $R=0.0005$ (squares);
$R=0.005$ (circles); $R=0.1$ (diamonds).}
\end{figure}

A systematic analysis of the filtering effect for different values
of $q$ is summarized in Figure~9 which shows the asymptotic value
for long times of the average fraction of cultural domains $g$ as
a function of $q$, with $F=5$, for different values of the
filtering probability $R$. When no filtering acts on the system
($R=0$) the behavior is that of the original Axelrod's model and
also coincides with the direct mass media models for $B=0$.

The effects of the filtering process in the culturally homogeneous
region, i.e., for parameter values $q < q_c$, is similar to that
of a direct influence of endogenous mass media. When the
probability $R$ is increased, the threshold value of $q$
decreases. There is a value $q_c(R)$ below which the system still
reaches a homogeneous cultural state under the influence of the
filter. An increase in $R$ for parameters $q < q_c(R)$ leads to
cultural diversity. Thus, both mechanisms of feedback information,
either direct or indirect, promote multiculturality in the region
of parameters where globalization prevails in the absence of any
feedback. The similar behavior found for the three types of mass
media influences considered here suggests that the phenomenon of
mass media-induced diversity should be robust in this region of
parameters, regardless of the type of feedback mechanism at work.

However, in the region of parameters $q > q_c$ where
multiculturality occurs for $R=0$ or $B=0$, the behavior of the
filter model differs from those of the direct mass media
influence. The filtering mechanism has little effect for values of
the probability $R<1$. As $R \rightarrow 1$ there is a small
decrease in the number of cultural groups formed in the system.
But at $R=1$ a discontinuity appears: the fraction of cultural
groups $g$ jumps from a value close to the one for $R=0$ to a
value close to $g=1$ corresponding to maximum multiculturality
(number of cultural groups equal to the number of agents in the
system). The case $R=1$ corresponds to an extreme restriction on
the dynamics, when no adoption of cultural features from neighbors
is allowed unless the state of the neighbor coincides with the one
of the global mass media. Since we are considering random initial
conditions, when $q$ is large enough, the probability that the
features of any agent coincide with those of the global mass media
message $M$ is quite small, making the convergence to
globalization, i.e., a common state with the media, very unlikely.
As a consequence, the random multicultural state subsists in the
system and manifests itself as a maximum value of $g$. The small
probability of interaction with the global mass media for large
values of $q$ when $R=1$ is also reflected in the very long
convergence time needed to reach the final multicultural state as
compared with the convergence time for $R<1$.

We note that the case $B=1$ in the model of direct global mass
media influence is less restrictive than the condition $R=1$ in
the filter model. Although local agent-agent interactions produce
negligible effects in both cases, in the model of direct influence, an
agent can still interact with the global mass media when there is
some cultural overlap between the agent and the mass media
message.

\begin{figure}[h]
\centerline{
\includegraphics[width=0.6\textwidth]{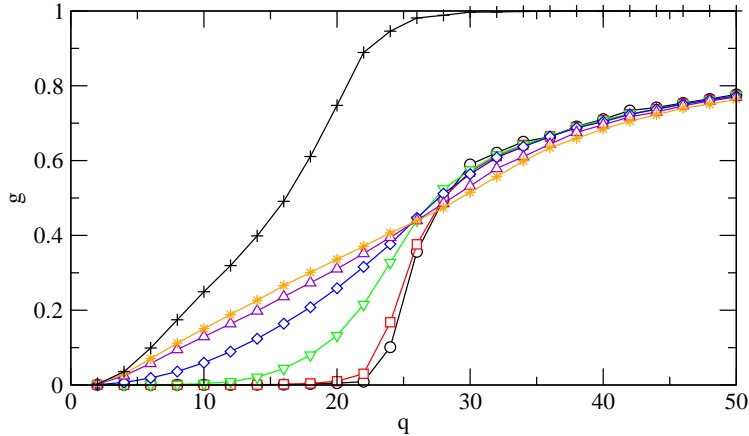}}
\caption{Average fraction of cultural domains $g$ as a function of
$q$, for different values of the probability $R$ for the filter
model. $R=0$ (circles); $R=0.01$ (squares); $R=0.1$ (triangles
down); $R=0.5$ (diamonds); $R=0.9$  (triangles up); $R=0.99$
(stars); $R=1.0$ (plus signs)}
\end{figure}

\section{Conclusions}

We have studied the effects of different forms of information
feedback on a social system in the framework of Axelrod's model
for cultural dissemination, including a model of local mass media influence. Two basic mechanisms of information
feedback have been interpreted as models of direct or indirect
mass media influence on the system \cite{Shibanai}. For the direct
information feedback mechanism we have considered the cases of
global mass media and local mass media influences \cite{PRE}. The
dynamics of direct interaction with the mass media is based on the
similarity or overlap between vector states. Although their origin
is different, at the local level the direct global and local mass
media influences act in the same manner, as a ``fifth" effective
neighbor whose specific source is not essential. Consequently they
produce similar effects in the system. First, we find the
nontrivial result that direct mass media can enhance cultural
diversity, in agreement with the results of \cite{Shibanai}: for
parameter values for which the system reaches a global culture due
to the local interaction among the agents, there is a threshold
value of the probability of interaction $B$ with mass media. For
values of $B$ above this threshold cultural diversity emerges.
This happens because there is a competition between the
consequences of the similarity rule applied to the agent-agent
interactions, and applied to the agent-mass media interaction.
This competition leads to the formation of cultural domains and to
polarized states. Secondly, we find another surprising effect at variance with the conclusions of \cite{Shibanai}: for
parameter values for which the dynamics based on the local
interaction among the agents produces a frozen multicultural
configuration, very weak directly interacting mass media produces
cultural homogeneity. This range of parameters was not explored in
Ref. \cite{Shibanai}, and we find here that mass media interaction
for $q>q_c$ produces the opposite effect of the one described
before. That is, it favors cultural homogeneization.The limit $B
\rightarrow 0$ is discontinuous and the homogenizing effect for
$B<<1$ occurs because the interaction with the mass media acts as
a perturbation on the non stable multicultural configurations with
frozen dynamics that appear for $B=0$. As a general result for any
value of $q$, we find that for small values of $B$, the
interaction with mass media promotes cultural homogenization: for
$q<q_c$ this interaction preserves homogeneity, while for $q>q_c$
it causes a drop in the degree of cultural diversity in the system
as measured by the average fraction of cultural domain $g$. We
have calculated, for the two types of direct mass media influences
considered, the corresponding boundary in the space of parameters
$(B,q)$ that separates globalization and polarization states. The
spatially nonuniform local mass media has a greater ordering
effect than the uniform global mass media in the regime $q>q_c$.
The range of values of $B$ for which globalization occurs for
$q<q_c$ is also larger for the nonuniform local mass media.

The effect of the indirect global mass media influence as a
filtering process in the culturally homogeneous region, i.e., for
$q < q_c$, is similar to that caused by a direct influence of mass
media. For small values of the filtering probability $R$ the
system reaches a culturally homogeneous state. For values of $R$
greater than a threshold value the system converges to a state of
cultural diversity. Thus, both mechanisms of feedback information,
either direct or indirect, promote multiculturality in a region of
parameters where it would not be present in the absence of any
feedback. In the region of parameters $q > q_c$ where
multiculturality occurs for either $B=0$ or $R=0$, the filtering
mechanism has, for values of the probability $R<1$, a very weak
effect in comparison to the one caused by a direct mass media
influences: there is only a small decrease in the number of
cultural groups formed. However, when the extreme restriction
$R=1$ is imposed, the number of cultural groups jumps
discontinuously to a value corresponding to maximum
multiculturality.

The similarity of the behaviors of the different types of mass
media influences considered here suggests that the phenomenon of
mass media-induced diversity should be robust, regardless of the
type of feedback mechanism. In spite of the differences mentioned
between direct and indirect information feedback processes as well
as between uniform and nonuniform mass media influences, it is
remarkable that the collective behavior of the agents displays
analogous phenomenology for the three types of mass media
considered.

Generally speaking, our analysis unveils the delicate compromise
between direct agent-agent interactions and feedback processes.
Mass media reflects local or global cultural trends created by
local agent-agent interaction, but mass media information is
processed by agent interactions, while the agent-mass media
interaction is conditioned by the overlap of the cultural features
of the agent and the mass media message. We have analyzed the
effect of different forms of mass media for the full range of the
parameter $q$ that measure an initial cultural diversity. Our
results indicate qualitatively different effects when
globalization $(q<q_c)$ or polarization $(q>q_c)$ would prevail
when no mass media feedback is taken in account. In summary, we
find that, when the probability of interacting with the mass media
is sufficiently large, mass media actually contribute to cultural
diversity in a social system, independently of the nature of the
media. But direct mass media influences are found to be efficient
in promoting cultural homogeneity in conditions of weak broadcast
of a message, so that local interactions among individuals can be
still effective in constructing some cultural overlap with the
mass media message. Strong media messages do not lead to cultural
homogenization because agent-agent interaction becomes
inefficient. These results identify the power of being subtle in
mass media massages. In addition, direct local mass media appear
to be more effective in promoting uniformity in comparison to
direct global broadcasts, which identifies the importance of local
media (feedback at regional levels) in the cultural globalization
path.

\section{Appendix}

 Color code for visualization of patterns. A different color can be assigned to each cultural state. Cultural domains correspond to sets of elements that share the same color. Figure 10 illustrates
 the case for $F=3$ and $q=2$.  In general for $q>2$, $q$ weights the basic colours (R,G,B): $0 \leq \sigma_{if} / (q-1)\leq1$.

\begin{figure}[h]
\centerline{
\includegraphics[width=0.6\textwidth]{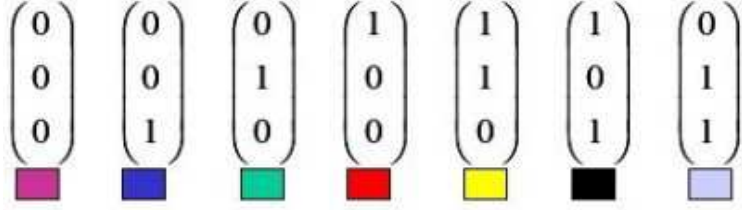}}
\caption{ For $F = 3$ and $q=2$ there are $q^F=8$ different states that can be assigned different color as shown.}
\end{figure}

\begin{center}
% use packages: array
\begin{tabular}{ll}

Key to formule employed.\\

$ l(i,j) = \sum_{f=1}^{F} \delta_{\sigma_{if},\sigma_{jf}}$   &  \small{Number of shared  features between agent $i$ and agent $j$}
\vspace{0.5cm} \\ 

$ p_{ij} = l(i,j)/F $                                         & \small{The probability that the agent i interact with the agent $j$}
\vspace{0.5cm}\\
$ B \cdot p_{iM}=B \cdot \frac{l(i,M)}{F} $                   &  \small{The total probablity that the agent i interact with the mass media}. 
\vspace{0.5cm}\\ 

  $ l(i,M) = \sum_{f=1}^{F} \delta_{\sigma_{if},\sigma_{Mf}}$   & \small{The overlap between agent $i$ and the mass media messages} 
\vspace{0.5cm}\\ 

 $ g = \frac{\langle N_{g} \rangle}{N} $                           & \small{The average fraction oif cultural domain}.
\vspace{0.5cm}\\ 

$ \delta_{\sigma{if},\sigma_{jf}}\left\{
      \begin{array}{c}
        if \; \; \sigma_{if}= \sigma_{jf}; \; \; \delta_{\sigma{if},\sigma_{jf}} = 1\\
        if \; \; \sigma_{if} \neq \sigma_{jf}; \; \; \delta_{\sigma{if},\sigma_{jf}} = 0
      \end{array}. \right.
$    & Kronecker delta definitio
\\

\end{tabular}
\end{center}

\section*{Acknowledgments}
J.\ C. \ G-A., V.\ M.\ E. and M.\ SM acknowledge financial support
from MEC (Spain) through project CONOCE2 (FIS2004-00953). M. G. C.
acknowledges support from FONACIT (Venezuela) under grant
F-2002000426. K K.\ K.\ acknowledges support from DFG
Bioinformatics Initiative BIZ-6/1-2 and from Deutscher
Akademischer Austausch Dienst (DAAD).

\end{document}